\documentclass[a4paper,11pt]{article}
\usepackage{pos}
\usepackage{wrapfig}
\usepackage{multicol}
\usepackage{slashed}
\usepackage{physics}
\usepackage{subcaption}

\newcommand{\cs}[1]{\mathcal{#1}}
\newcommand{\MUB}{\hat{\mu}_{\mathcal{B}}}
\newcommand{\MUQ}{\hat{\mu}_{\mathcal{Q}}}
\newcommand{\MUS}{\hat{\mu}_{\mathcal{S}}}

\newcommand{\BB}{\chi_{\mathcal{BB}}}
\newcommand{\QQ}{\chi_{\mathcal{QQ}}}
\renewcommand{\SS}{\chi_{\mathcal{SS}}}
\newcommand{\BQ}{\chi_{\mathcal{BQ}}}
\newcommand{\BS}{\chi_{\mathcal{BS}}}
\newcommand{\QS}{\chi_{\mathcal{QS}}}

\title{QCD equation of state in the presence of magnetic fields at low density}
\author[a]{S. Bors\'anyi}
\author[b]{B. B. Brandt}
\author[b]{G. Endr\H{o}di}
\author[a]{J. Guenther}
\author[a]{R. Kara}
\author*[b]{A. D. M. Valois}

\affiliation[a]{University of Wuppertal, Department of Physics, Wuppertal D-42119, Germany}

\affiliation[b]{Universität Bielefeld, Universitätsstraße 25, 33615 Bielefeld, Germany}

\emailAdd{borsanyi@uni-wuppertal.de}
\emailAdd{brandt@physik.uni-bielefeld.de}
\emailAdd{endrodi@physik.uni-bielefeld.de}
\emailAdd{jguenther@uni-wuppertal.de}
\emailAdd{rkara@uni-wuppertal.de}
\emailAdd{dvalois@physik.uni-bielefeld.de}

\newcommand{\ave}[1]{\left\langle\hspace{0.1cm}#1\hspace{0.1cm}\right\rangle}

\abstract{Peripheral heavy-ion collisions are expected to exhibit magnetic fields with magnitudes comparable to the QCD scale, as well as non-zero baryon densities. Whereas QCD at finite magnetic fields can be simulated directly with standard lattice algorithms, the implementation of real chemical potentials is hindered by the infamous sign problem. Aiming to shed light on the QCD transition and on the equation of state in that regime, we carry out lattice QCD simulations with 2+1+1 flavors of staggered quarks with physical masses at finite magnetic fields and employ a Taylor expansion scheme to circumvent the sign problem. We present the leading-order coefficient of the expansion calculated at non-zero magnetic fields and discuss the impact of the field on the strangeness neutrality condition.}
\FullConference{%
 The 40th International Symposium on Lattice Field Theory, LATTICE2023
  July 31 to August 4, 2023
  Fermilab, Illinois
}
\begin{document}
\maketitle

\section{Introduction}\label{sec:intro}
The impact of high baryon densities on strongly interacting matter is relevant for the phenomenology of future heavy-ion collision experiments, such as FAIR and NICA, as well as for the description of the inner core of compact neutron stars. However, simulating the underlying theory, QCD at high baryon densities, using lattice techniques remains a formidable challenge and subject of ongoing research (for a recent review, see \cite{Nagata:2021ugx}). The biggest caveat is the emergence of a sign problem and its increasing severity at high density, which hinder the exploration of the QCD phase diagram deep into the low temperature ($T$) and high baryon chemical potential ($\mu_{\cs{B}}$) region, necessary for the description of the above-mentioned systems. Therefore, due to the prohibitive nature of the sign problem, alternative techniques are necessary to circumvent it, such as reweighting, Taylor expansion, or analytical continuation from imaginary chemical potentials. However, these techniques are restricted to low baryon densities and hence a thorough understanding of hot and dense QCD from first principles remains elusive. Furthermore, an ongoing quest for the conjectured critical endpoint in the phase diagram, where the QCD crossover would become a first-order transition, has received increasing efforts in the lattice community (for a review, see \cite{philipsen2007lattice}), as well as
via functional methods, see e.g. the recent review~\cite{Dupuis:2020fhh}.

\begin{figure}[!h]
    \centering
    \includegraphics[width=0.6\textwidth]{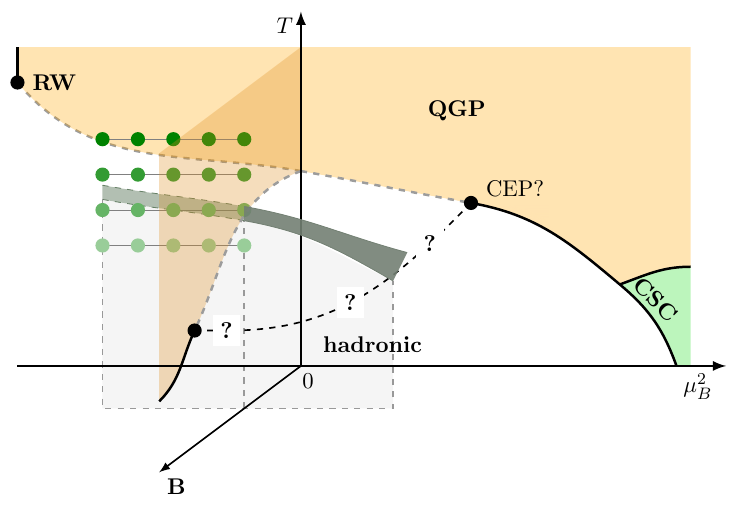}
    \caption{Conjectured three-dimensional QCD phase diagram in the temperature -- magnetic field -- baryon chemical potential space. The left side of the figure corresponds to imaginary, while the right side to real baryon chemical potentials. The green dots indicate our current simulations and the gray band is a possible outcome for the transition temperature. }
    \label{fig:phase_diagram}
\end{figure}

Additionally, strong magnetic fields are also expected in peripheral heavy-ion collisions and certain types of neutron stars (magnetars). In contrast to a baryon chemical potential, magnetic fields are free of a sign problem and can be directly simulated using lattice QCD. Moreover, the magnetic field is known to strengthen the QCD crossover~\cite{Endrodi:2015oba}. At large enough $B$, it was argued that the nature of the deconfinement transition in QCD changes to first order \cite{cohen2014new}, which was later supported by lattice QCD studies of the phase diagram, with a predicted critical endpoint in the range $4\text{ - }9 \text{ GeV}^2$ \cite{d2022phase}. It is tempting to conjecture that the critical endpoints in the $T$-$\mu_{\cs{B}}$ and $T$-$B$ planes might even be connected by a single critical line, raising interesting questions on the relationship between magnetic fields and baryon densities as probes of the QCD phase diagram (see also Refs.~\cite{braguta2019finite,Ding:2021cwv}). In Figure~\ref{fig:phase_diagram} we sketch the conjectured phase diagram in the $T$-$\mu_{\cs{B}}$-$B$ space.
In the context of heavy-ion collisions, the QCD equation of state at nonzero density is an important ingredient for model simulations and it has been computed for both zero \cite{Borsanyi:2012cr} and non-zero \cite{Kolomoyets:2021onx} magnetic fields. Also phenomenologically relevant for the heavy-ion setup is the strangeness neutrality condition. During the chemical freeze-out stage, the strangeness content of the fireball averages zero, and the parameters for which this happens can be determined from fluctuations of conserved charges \cite{Bazavov:2012vg}.
This property has been previously studied on the lattice for a vanishing magnetic field \cite{Guenther:2023aeb}. In this work, we extend our previous studies and investigate, for the first time, the joint impact of a chemical potential and a magnetic field on the strangeness neutral point using a Taylor expansion framework. We also compute the equation of state in the presence of $\mu_{\cs{B}}$ and $B$ to shed light on their role in QCD thermodynamics.
While here we assume the magnetic field to be constant in space, 
the case of inhomogeneous magnetic fields -- expected to be more 
relevant for the heavy-ion collision setting -- can also be 
simulated directly on the lattice~\cite{Brandt:2023dir}, and the impact of a non-zero chemical potential included in the same way. 

This proceedings article is organized as follows: in Section~\ref{sec:mag_lat} we describe our magnetic field background and its implementation on the lattice. This is followed by Section~\ref{sec:setup}, where we give the details of our simulation setup. In Section~\ref{sec:strangeness} we discuss our results on the impact of a non-zero magnetic field on the strangeness neutrality condition. In Section~\ref{sec:eos} we present our determination of the leading-order (LO) coefficient in the Taylor expansion of the pressure for three magnetic field strengths, and in Section~\ref{sec:conclusions} we summarize our findings and state our conclusions.
\section{Background magnetic field on the lattice}\label{sec:mag_lat}
In a finite periodic volume, the magnetic field flux is quantized. On the lattice, the presence of such a field is implemented by multiplying the $\mathrm{SU}(3)$ links by complex phases associated with the magnetic field. For multiple flavors, these phases are defined as $u^f_{\mu} = e^{iaq_fA_{\mu}}$, where $q_f$ denotes the charge of the quark flavor $f$ and $A_{\mu}$ is the $\mathrm{U}(1)$ background gauge field. For convenience, $A_{\mu}$ is defined so that the phases satisfy periodic boundary conditions. In particular, in the case of a uniform magnetic field pointing in the $z$ direction the links are as follows \cite{bali2012qcd}
\begin{align}
u^f_{x}(x,y,z,t) &= 
    \left\{
        \begin{array}{ll}
        e^{-iq_fBL_xy} & \mbox{if } x = L_x-a \nonumber\\
        1 & \mbox{if } x \neq L_x-a
        \end{array}
    \right. \\
u^f_{y}(x,y,z,t) &= e^{iaq_fBx} \hspace{4cm} \text{with} \hspace{0.5cm} B = \frac{2\pi N_b}{qL_xL_y},\hspace{1cm} N_b\in\mathbb{Z}.\\
u^f_z(x,y,z,t) &= 1 \nonumber\\
u^f_t(x,y,z,t) &= 1 \nonumber
\end{align}
where $q$ is a reference charge given by the modulus of the greatest common divisor of the quark charges, \textit{i.e.} for QCD $q = e/3$, such that $q_f/q \in \mathbb{Z}$, and $e$ is the elementary electric charge. Throughout this text, we express the magnetic field dependence of observables in terms of the renormalization-group-invariant quantity $eB$.
\section{Simulations setup and observables}\label{sec:setup}
We simulate 2+1+1 flavors of staggered quarks with masses at the physical point using the Symanzik improved gauge action with four stout smearing steps. We carry out the simulations on a $32^3\times 8$ lattice for three magnetic field strengths, namely $eB = 0$, $0.5$, $0.8\text{ GeV}^2$, and scan a range of temperatures from 125 to 300 MeV in the $B=0$ case, and from 135 to 200 MeV in the $B > 0$ cases.
We change the temperature by changing the lattice spacing $a$ via the inverse gauge coupling $\beta$.
This implies that for a given $N_b$, the physical value of $eB$ changes for different temperatures. Therefore, to achieve the desired values, for each $eB \neq 0$ we do simulations using two values of $N_b$, corresponding to the closest neighboring integers.

To circumvent the sign problem, we Taylor-expand the pressure (normalized by $T^4$) in the quark chemical potential basis as follows
\begin{equation}
\frac{P}{T^4} = \sum_{ijk}\chi_{ijk}\,\hat{\mu}_{u}^i\,\hat{\mu}_{d}^j\,\hat{\mu}_{s}^k\,, \hspace{0.5cm} \text{where} \hspace{0.5cm} \chi_{ijk} = \frac{1}{VT^3}\qty(\pdv{\hat{\mu}_u})^i\qty(\pdv{\hat{\mu}_d})^j\qty(\pdv{\hat{\mu}_s})^k\log Z\Big|_{\mu_u=\mu_d=\mu_s=0},
\label{eq:pressure_taylor}
\end{equation}
$\hat{\mu}_X \equiv \mu_X/T$, with $X = u,d,s$, and $Z$ is the grand canonical partition function. We compute the coefficients $\chi_{ijk}$ appearing in the sum above and relate them to the susceptibilities in the physical basis $\{\mu_{\cs{B}}, \mu_{\cs{Q}}, \mu_{\cs{S}}\}$ using the transformations
\begin{equation}
\mu_{u} = \frac{1}{3}\mu_{\cs{B}} + \frac{2}{3}\mu_{\cs{Q}}\,, \hspace{0.5cm} \mu_{d} = \frac{1}{3}\mu_{\cs{B}} - \frac{1}{3}\mu_{\cs{Q}}\,, \hspace{0.5cm} \mu_{s} = \frac{1}{3}\mu_{\cs{B}} - \frac{1}{3}\mu_{\cs{Q}}-\mu_{\cs{S}}\,.
\end{equation}
On the physical basis, the susceptibilities are related to the ones in the quark basis by
\begin{align}
    \BB &= \frac{1}{9}\qty(\chi_{200} + \chi_{020} + \chi_{002} + 2\chi_{110} + 2\chi_{101} + 2\chi_{011}) \\
    \QQ &= \frac{1}{9}\qty(4\chi_{200} + \chi_{020} + \chi_{002} - 4\chi_{110} - 4\chi_{101} + 2\chi_{011}) \\
    \SS &= \chi_{002}
\end{align}
Inserting these relations in Eq.~\eqref{eq:pressure_taylor} for the pressure we obtain to quadratic order in the chemical potentials,
\begin{equation}
\frac{P}{T^4} = c_0 + \frac{\BB}{2}\MUB^2 + \frac{\QQ}{2}\MUQ^2 + \frac{\SS}{2}\MUS^2 + \BQ\MUB\MUQ + \BS\MUB\MUS + \QS\MUQ\MUS.
\label{eq:P_T4}
\end{equation}
To make closer contact with the heavy-ion collision setup, Eq.~\eqref{eq:P_T4} can be further constrained in terms of the chemical potentials. In the next section, we discuss these constraints and compute the necessary and sufficient conditions for them to be satisfied.
\section{Strangeness neutrality and isospin asymmetry}\label{sec:strangeness}
\begin{figure}[!h]
    \centering
    \begin{subfigure}{0.49\textwidth}
    \includegraphics[width=\linewidth]{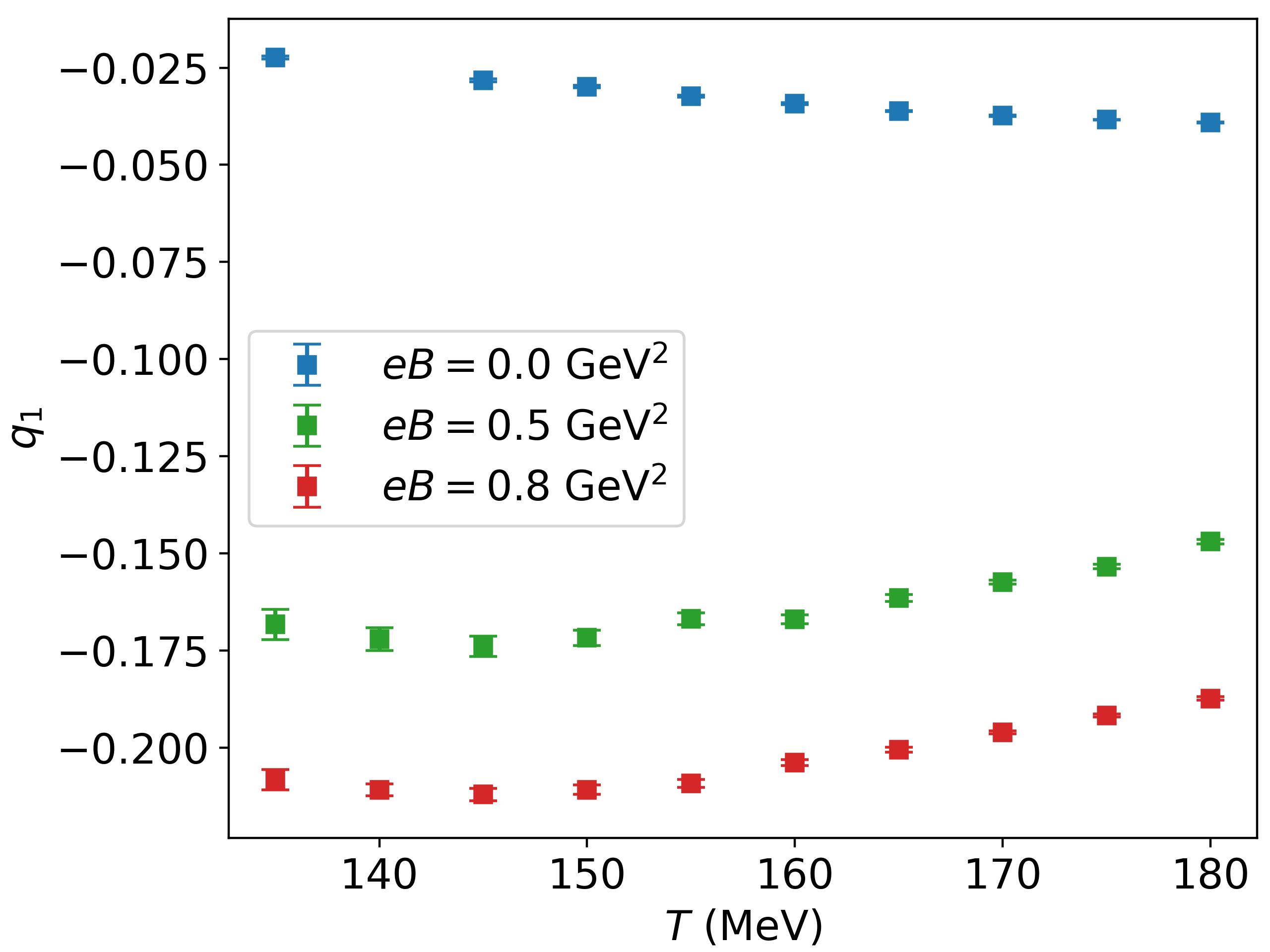}
    \end{subfigure}
    \begin{subfigure}{0.49\textwidth}
    \includegraphics[width=\linewidth]{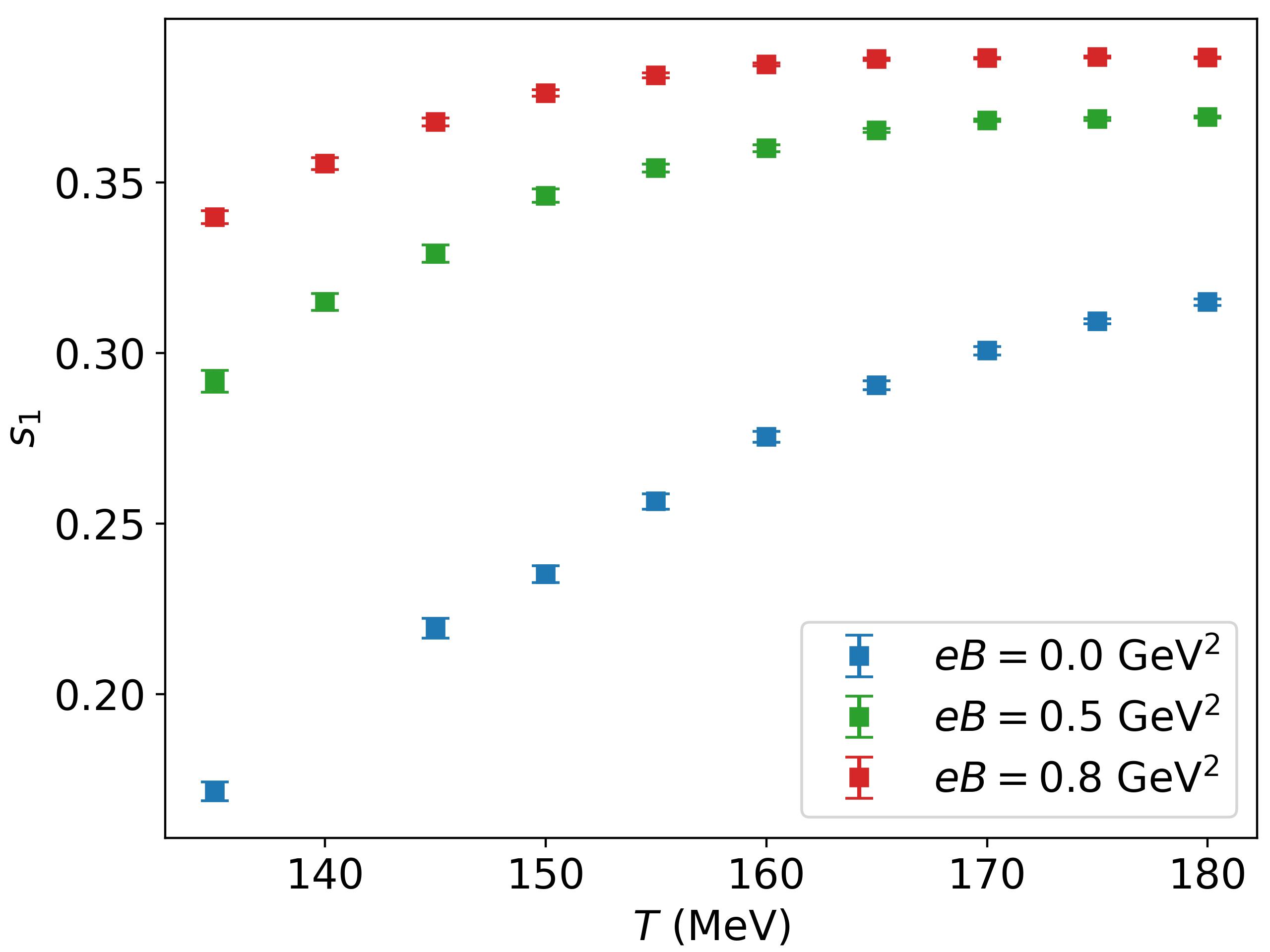}
    \end{subfigure}
    \caption{Linear coefficients of the charge (left) and strange (right) chemical potentials in terms of the baryon chemical potential for the strangeness neutrality condition as a function of $T$ for different magnetic fields.}
    \label{fig:q1_s1}
\end{figure}

In a heavy-ion collision experiment, the number of valence quarks present in the colliding ions constrains the conserved quantities in the baryon, charge, and strangeness basis. For Pb atoms, for instance, the fractions of $u$, $d$, and $s$ quarks are $53.5\%$, $46.5\%$ and $0\%$, respectively. Therefore, we expect strangeness neutrality $\ave{n_\cs{S}} = 0$, as well as $\ave{n_\cs{Q}}/\ave{n_\cs{B}} \approx 0.4$, which gives a slight deviation from the isospin-symmetric point ($\ave{n_\cs{Q}}/\ave{n_\cs{B}} = 0.5$).
To achieve these conditions in our lattice simulations, we express $\mu_{\cs{Q}}$ and $\mu_{\cs{S}}$ in terms of $\mu_{\cs{B}}$ perturbatively
\begin{equation}
\mu_{\cs{Q}} = q_1\mu_{\cs{B}} + \mathcal{O}(\mu_{\cs{B}}^3)\,, \hspace{1cm} \mu_S = s_1\mu_{\cs{B}} + \mathcal{O}(\mu_{\cs{B}}^3)\,,
\end{equation}
and determine the leading-order coefficients $q_1$ and $s_1$ such that the desired constraints are fulfilled. In terms of the susceptibilities of conserved charges, the coefficients are given by \cite{Bazavov:2012vg}
\begin{equation}
 q_1 = \frac{0.4\qty(\chi_{\cs{B}\cs{B}}\chi_{\cs{S}\cs{S}}-\chi_{\cs{B}\cs{S}}^2) -\qty(\chi_{\cs{B}\cs{Q}}\chi_{\cs{S}\cs{S}}-\chi_{\cs{B}\cs{S}}\chi_{\cs{Q}\cs{S}})}{\chi_{\cs{Q}\cs{Q}}\chi_{\cs{S}\cs{S}} - \chi_{\cs{Q}\cs{S}}^2-0.4\qty(\chi_{\cs{B}\cs{Q}}\chi_{\cs{S}\cs{S}} - \chi_{\cs{B}\cs{S}}\chi_{\cs{Q}\cs{S}})}\,,  \hspace{1cm} s_1 = -\frac{(\chi_{\cs{B}\cs{S}}+q_1\chi_{\cs{Q}\cs{S}})}{\chi_{\cs{S}\cs{S}}}.
\end{equation}
In Figure~\ref{fig:q1_s1} we show the coefficients $q_1$ and $s_1$ as a function of temperature for different magnetic fields.
We now discuss the physical interpretation of this behavior in terms of the hadronic content of the system, inspired by the hadron resonance gas (HRG) model.

In the isospin-symmetric case, and with $eB = 0$, we would have $\ave{n_Q}/\ave{n_B} = 0.5$, and the system would consist of an equal number of protons and neutrons. However, to have a ratio slightly below the isospin point, the positive charges have to be suppressed, producing an imbalance between protons and neutrons. This suppression is encoded in the fact that $q_1$ is negative, as shown on the left side of Figure~\ref{fig:q1_s1}. 
At $eB\neq0$, both protons and neutrons (more precisely, their spin states with lowest energy) become lighter~\cite{Endrodi:2019whh} due to their nontrivial magnetic moments. This enhances the effect further, and in order to keep the ratio $\ave{n_Q}/\ave{n_B}$ fixed, given the aforementioned $u$, $d$, and $s$ quark abundances, the protons have to be suppressed even more. Therefore, $q_1$ increases in magnitude and moves towards more negative values. We can formulate a similar argument for the behavior of $s_1$, on the right plot, in terms of the $\Lambda^0$ baryon (the lightest hadron with non-zero strangeness, which also features a nonzero magnetic moment), but since the strange quark has negative strangeness by definition, the behavior is the opposite and $s_1$ is enhanced by the magnetic field.
Next, we apply these coefficients for the computation of the equation of state in leading order.
\section{Leading-order Taylor coefficient}\label{sec:eos}
\begin{figure}[!h]
    \centering
    \includegraphics[width=\textwidth]{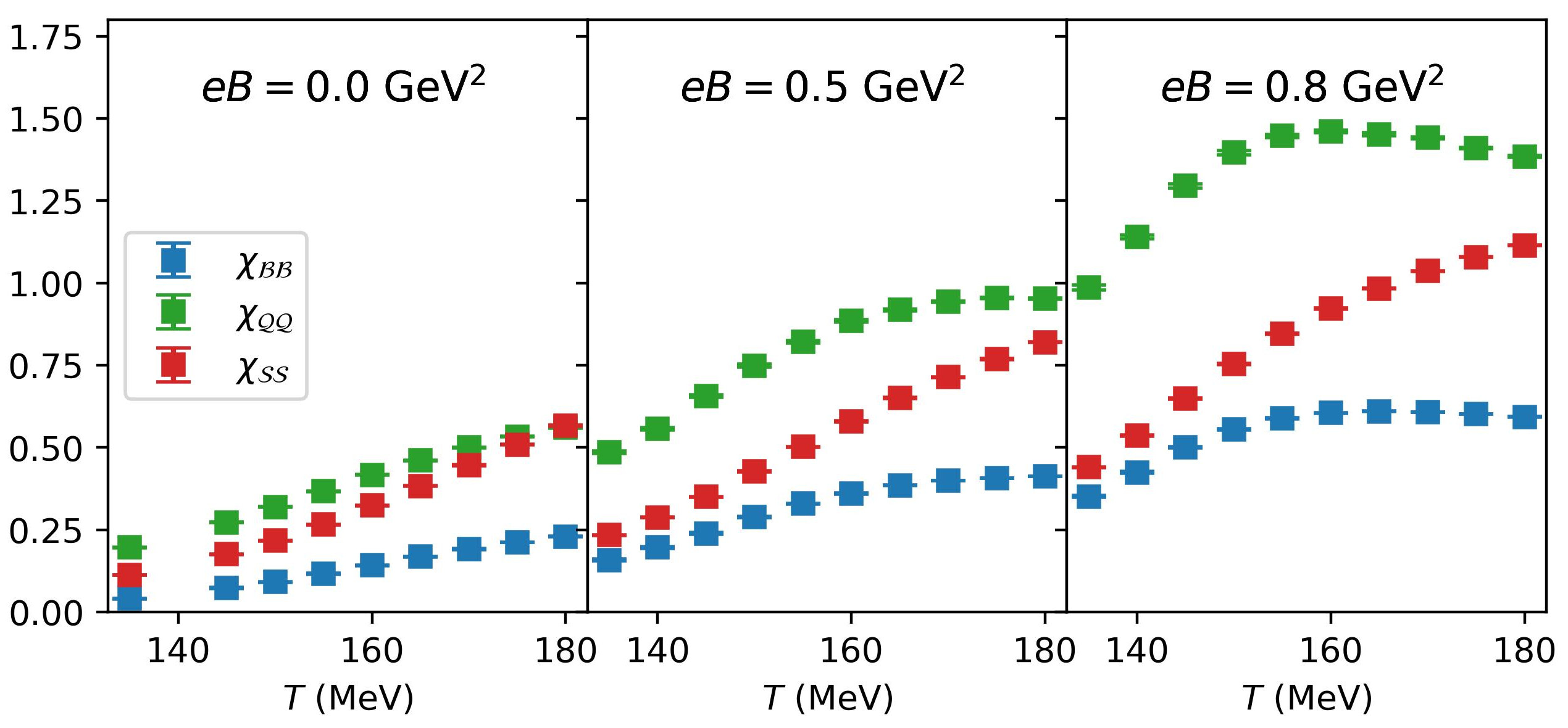}
    \caption{$\BB$, $\QQ$ and $\SS$ susceptibilities as functions of $T$ for different magnetic field strengths.}
    \label{fig:BB_QQ_SS}
\end{figure}

\begin{figure}[!h]
    \centering
    \begin{subfigure}{0.49\textwidth}
    \includegraphics[width=\linewidth]{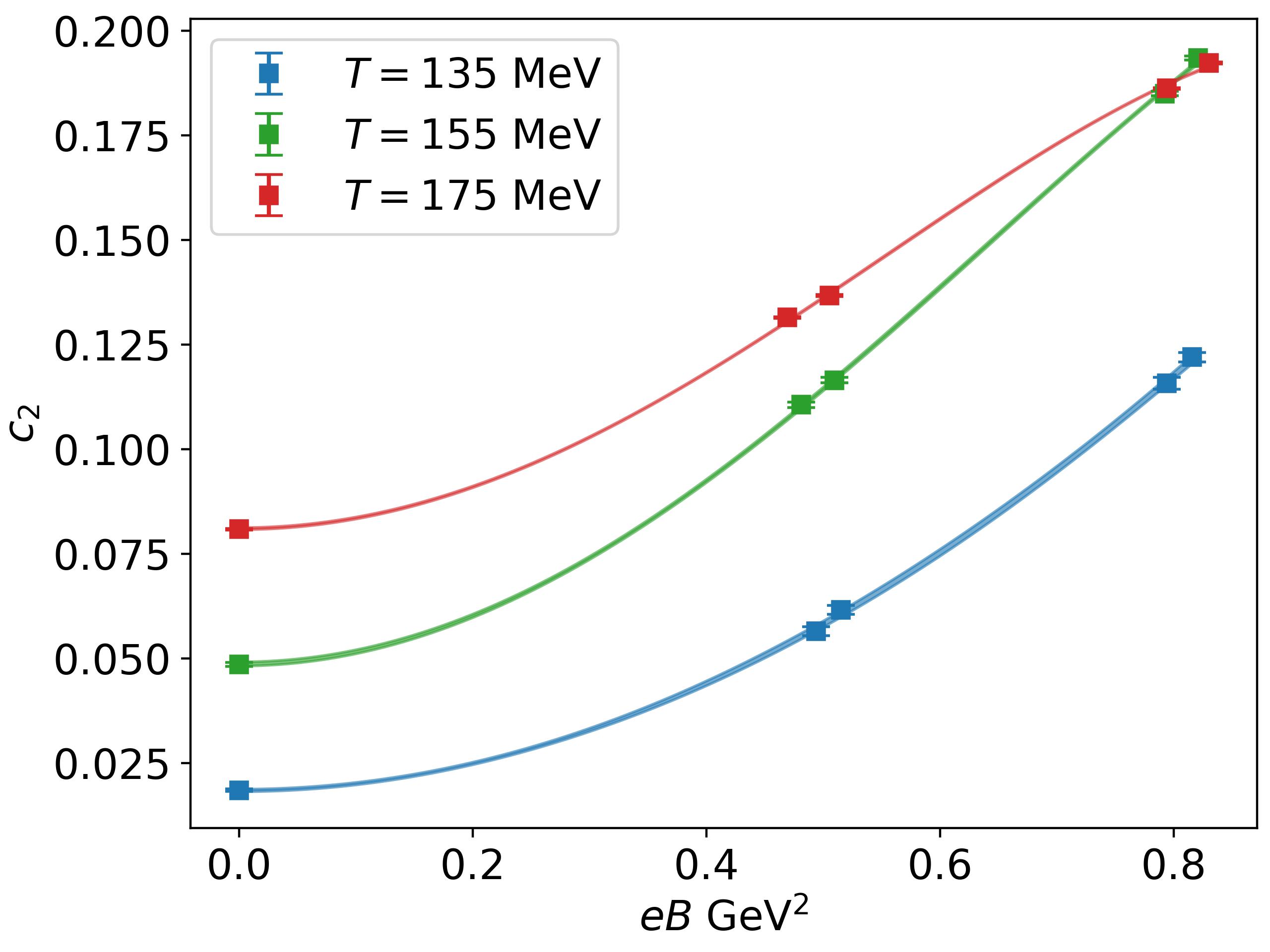}
    \end{subfigure}
    \begin{subfigure}{0.49\textwidth}
    \includegraphics[width=\linewidth]{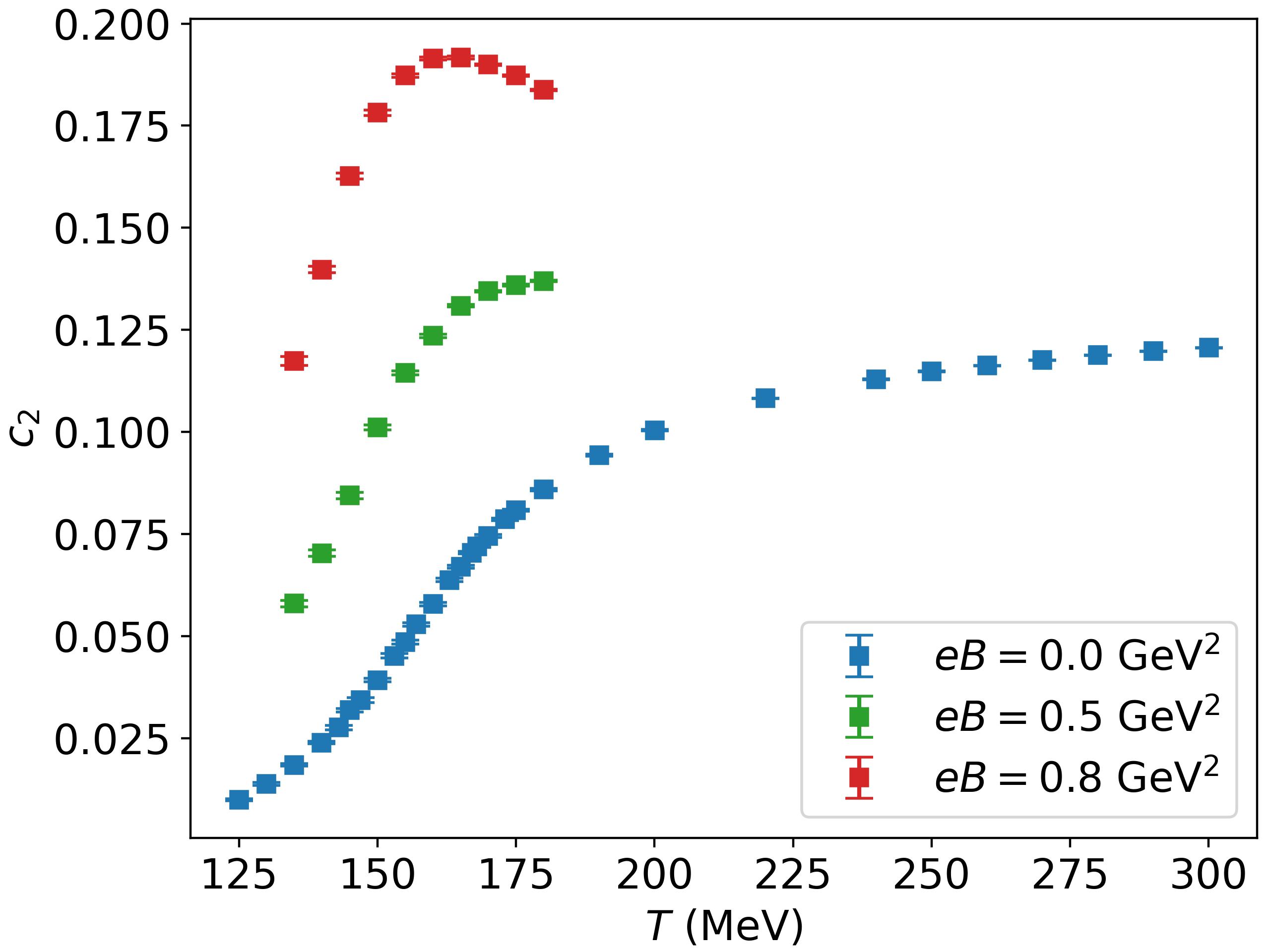}
    \end{subfigure}
    \caption{Left plot: leading-order contribution to the equation of state, $c_2$, as a function of the magnetic field for different temperatures. Right: $c_2$ as a function of temperature for different magnetic fields.}
    \label{fig:c2_B_T}
\end{figure}

Using the computation of the susceptibilities and the coefficients $q_1$ and $s_1$ done in the previous section, we can express the pressure in terms of the baryon chemical potential
\begin{equation}
\frac{P}{T^4} = c_0 + \qty(\frac{1}{2}\BB + \frac{1}{2}\QQ q_1^2 + \frac{1}{2}\SS s_1^2 + \BQ q_1 + \BS s_1 + \QS q_1s_1)\hat{\mu}_{\cs{B}}^2\,.
\end{equation}
The coefficient $c_0$ represents the pressure at zero density, whose
$B$-dependence is characterized for weak magnetic fields by 
the magnetic susceptibility of QCD matter (see e.g.~Ref.~\cite{Bali:2020bcn}).
In turn, the coefficient of $\hat{\mu}_{\cs{B}}^2$ is the leading-order contribution to the pressure due to nonzero density and we define it as $c_2$. In Figure \ref{fig:BB_QQ_SS} we show three susceptibilities that contribute to $c_2$. 
In Figure~\ref{fig:c2_B_T}, we show our results for the behavior of this coefficient as a function of $T$ and $B$. Due to the parity symmetry of the partition function, $Z(B) = Z(-B)$, the $c_2$ coefficient can only contain even powers of the magnetic field. Therefore, we fit our data using a quartic function to account for the non-quadratic behavior at high $B$. If we associate the inflection point of $c_2$ with the pseudo-critical temperature ($T_c$) of the QCD crossover, we observe that the transition is shifted to lower values of $T_c$ as the magnetic field grows, as expected from the known behavior of other observables.
\section{Summary and Outlook}\label{sec:conclusions}
In this work, we carried out lattice simulations to study the QCD transition and the equation of state in the presence of both magnetic fields as well as baryon chemical potentials, employing a Taylor expansion framework for the latter. We determined, for the first time in the literature, the impact of the magnetic field on the strangeness neutral line in the presence of an isospin imbalance at low densities by computing the linear coefficients $q_1$ and $s_1$ as functions of $eB$ and $T$. We also provided a physical picture of the behavior of these coefficients with the magnetic field in terms of the hadronic content of the system. We observed that $q_1$ is suppressed to negative values, while $s_1$ is enhanced by the magnetic field, and noted that this can be explained by the suppression of protons and $\Lambda^0$ baryons. Using the strangeness neutral line defined by these coefficients, we also computed the leading-order contribution to the equation of state for several values of the magnetic field.

In the future, we aim to extend this study to higher-order susceptibilities and compute the equation of state at higher $\mu_B$, as well as using finer lattices and finally computing the continuum limit of our observables. \\

\noindent
{\bf Acknowledgments}\\
This research has received funding from the programme "Netzwerke 2021", an initiative of the Ministry of Culture and Science of the State of Northrhine Westphalia.
The authors gratefully acknowledge the Gauss Centre for Supercomputing e.V.
(\href{https://www.gauss-centre.eu}{\tt www.gauss-centre.eu})
for funding this project by providing computing time on the GCS Supercomputer
SuperMUC-NG at Leibniz Supercomputing Centre
(\href{https://www.lrz.de}{\tt www.lrz.de}).
\bibliographystyle{utphys}
\bibliography{bibliography.bib}
\end{document}